\documentclass[prc,aps,showpacs,twocolumn,superscriptaddress,nofootinbib]{
revtex4-1}
\usepackage{graphicx} 
\usepackage{multirow} 
\usepackage{amsmath}
\usepackage{bm}

\begin{document}

\title{High-$x$ structure function of the  virtually free neutron}

\author{Wim Cosyn}
\affiliation{Ghent University, 9000 Ghent, Belgium}
\email{wim.cosyn@ugent.be}
\author{Misak~M.~Sargsian}
\affiliation{Florida International University, Miami, Florida 33199, USA}
\email{sargsian@fiu.edu}

\date{\today}

\begin{abstract} 
The pole extrapolation method is applied  to the 
semi-inclusive  inelastic electron scattering off the deuteron with tagged 
spectator protons 
to extract the high-$x$  structure function of the neutron. This approach is 
based on the 
extrapolation of the measured cross sections at different momenta of 
the spectator proton   to the  non-physical pole of the  bound neutron 
in the deuteron.   The advantage of the method is 
in the possibility 
of suppression of the nuclear effects in a maximally model-independent way. 
The neutron structure functions obtained in this way demonstrate 
a surprising $x$ dependence at $x\ge 0.6$ and $1.6 \leq Q^2 \leq 3.38$ GeV$^2$, 
indicating a possible 
rise of the neutron to proton structure function ratio. 
If  the observed rise is valid in the true deep inelastic region  then it  
may 
indicate new dynamics in the generation of high-$x$ quarks in 
the nucleon.  One such mechanism we discuss is the possible dominance of 
 short-range isosinglet quark-quark correlations that can enhance the 
$d$-quark  distribution in the proton.
\end{abstract}

\maketitle

\section{Introduction}
\label{sec1}
Detailed knowledge of the $u$- and $d$- quark  
densities at large Bjorken $x$ is 
one of the important unresolved issues 
in  the QCD structure of  the nucleon.  This structure is very sensitive to 
quark correlation  dynamics at 
short distances~\cite{Feynman}.
The  high-$x$  distribution  is  important also to  
Large Hadron Collider (LHC) physics, in which due to 
QCD evolution,   the partons   at very large virtualities are 
sensitive to the high-$x$ quark distributions measured at lower $Q^2$.

The extraction  of the separate $u$- and $d$-quark distributions in the 
nucleon requires either the  measurement  of  the deep inelastic scattering 
(DIS) 
structure function of  the
proton and neutron or  weak interaction measurements off the proton in 
the charged current sector.  
Currently, the bulk of the data comes from the studies of inclusive DIS off the  
proton  and deuteron, with 
the latter being used  to extract  the neutron structure functions.  
In this case, 
nuclear effects such as the 
relativistic motion of the bound nucleons and their medium modification in the deuteron  become 
increasingly important at higher $x$,
rendering the 
extracted neutron structure functions strongly model dependent 
(see, e.g., Refs.~\cite{FS88,MT,Arrington}). 

One 
solution to the problem is to consider a new generation of 
experiments  in which DIS 
off the deuteron is followed by the detection of a recoil proton 
~\cite{MSS97,hnm}, i.e,
\begin{equation}
e + d \rightarrow e^\prime + X + p \,.
\label{semidis}
\end{equation}
 Such processes are
more complex due to 
the large final-state interactions~(FSI) of the DIS products with the 
spectator proton at large  momenta~\cite{CS11,Ciofi11}. 
Their advantage  however lies in the possibility of applying 
the pole extrapolation  procedure~\cite{polext06,polext11} at small momenta 
of 
the proton, in which case all 
nuclear effects due to Fermi motion, FSI and medium modification can be significantly suppressed in 
a practically model-independent way.

In Sec.~\ref{sec2} we introduce the general concept of pole extrapolation and explain why it is best suited for reactions involving 
deuteron target.  Sec.~\ref{sec3} presents the theoretical framework of tagged deep inelastic scattering which is then used in 
Sec.~\ref{sec4} to elaborate the pole extrapolation procedure for reaction~(\ref{semidis}).  In Sec.~\ref{sec5} we present the details 
of the pole extrapolation applied to the recent Barely Offshell Nucleon 
Scattering (BONuS) data and our results for neutron structure function at large 
Bjorken x.
In Secs.~\ref{sec6} and \ref{sec7} we discuss the results and present the  
conclusions. 

\section{General Concept of Pole Extrapolation}
\label{sec2}

The pole extrapolation was 
first suggested by Chew and Low~\cite{ChewLow} 
for probing the structure of so-called free $\pi$-mesons or the neutron by 
studying (a) $h+p\rightarrow h^\prime+ \pi + N_s$ and (b) $h+p\rightarrow 
h^\prime+  n + \pi^+_s$  reactions.  In these reactions, $N_s$ 
and $\pi^+_s$ can be considered as spectators to the underlying $h+\pi\rightarrow 
h^\prime + \pi$ and $h + p \rightarrow h^\prime +  n$ subprocesses, 
in which $h$ is an external probe.
Their idea was that by extrapolating the invariant momentum transfer to the 
unphysical pole 
values of the bound particles ($m_\pi$ and $m_n$ in this case), it will 
be possible to extract the free cross sections of the underlying 
subprocesses.  

The general concept of pole extrapolation can be seen if one considers a
target $A$ that consists of two bound constituents $B$ and $C$ 
in the reaction in which $B$ is probed by a 
particle $h$, while the 
particle $C$ emerges as a spectator.  
\begin{figure}[ht]
\centering\includegraphics[scale=0.4]{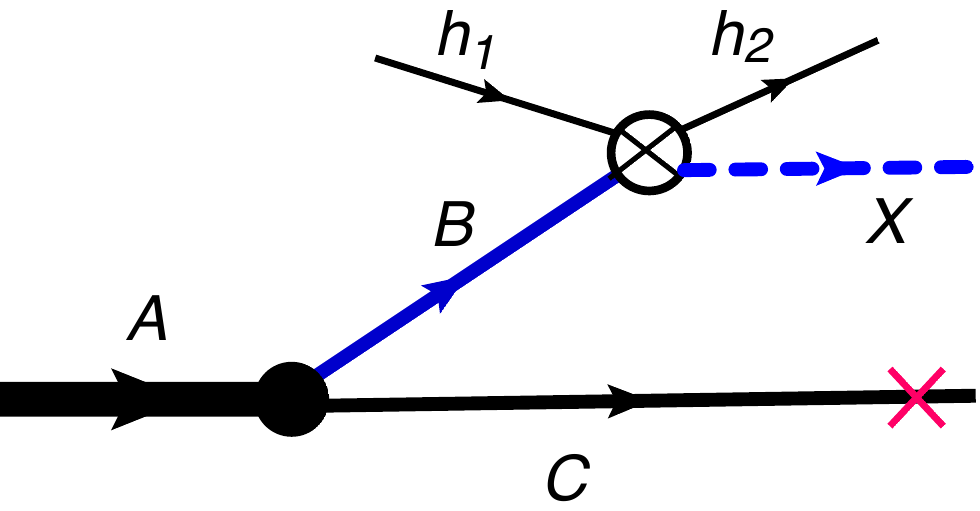}
\caption{(Color online) IA contribution to $ h_1+ A \rightarrow h_2 + C + 
X$ reaction, with $C$ acting as a spectator.}
\label{IAterm}
\end{figure}
The impulse approximation~(IA) (Fig.~\ref{IAterm})
amplitude of  such process has a structure
\begin{equation}
M_{IA} = M^{h_1+B\rightarrow h_2 + X}\frac{G(B)}{t - M_B^2}  
\chi^\dagger_C\Gamma^{A\rightarrow BC}\chi_A,
\label{IA}
\end{equation}
where $\chi_A$ and $\chi_C$ represent the wave functions of incoming composite 
particle $A$ and  outgoing spectator particle $C$.  The vertex 
$\Gamma^{A\rightarrow BC}$ characterizes the $A\rightarrow BC$ 
transition and the propagator of 
bound particle $B$  is described by $\frac{G(B)}{t - M_B^2}$, with $t = (p_A 
- p_C)^2$. 
As it follows from Eq.~(\ref{IA}), the IA amplitude has a singularity 
in the non-physical limit $t\rightarrow M_B^2$.  The most important property 
which makes this singularity significant is the 
so-called loop theorem~\cite{polext06}, according to which any other 
process beyond IA will  not  be singular due to a loop integration. 
Thus, even though non-IA terms can be large 
in the physical domain, they will be 
corrections in the $t\rightarrow M_B^2$ limit.  

The accuracy of the extrapolation depends  on the magnitude of 
\begin{equation}
l = m_B^2 - t_{\text{thr}}\,,
\label{extdist}
\end{equation}
where $t_{\text{thr}}$ is
the threshold value for the physical 
domain.

For the reaction (a) $t_{\text{thr}} = 0$, 
$l = m_\pi^2\approx 0.02$~GeV$^2$, and for  
(b) $ t_{\text{thr}} = 
(m_N - m_\pi)^2$, $l = 2m_Nm_\pi - m_\pi^2\approx 0.24$~GeV$^2$.  
  While $l$ is small for  reaction (a), the 
  problem is that the pole is positive and 
$t<0$,  so the  extrapolation 
requires a  crossing of the $t=0$ point which makes the result very sensitive 
to small variations  
in the method of extrapolation.  For reaction (b), even if 
one stays in  positive domain of $t$, $l$ is quite large, introducing 
ambiguities in the analytic form of the pole extrapolation.

It was observed in Ref.~\cite{polext06} that pole extrapolation is well suited 
for reactions  (\ref{semidis}) for which 
$A\equiv d$,  $B\equiv n$ and $C\equiv p$.  
In this case $t_{\text{thr}} = (M_d - m_p)^2$ and the variable $l$
is very small,  $l = 2m_n |\epsilon_b| - \epsilon_b^2\approx 0.004$~GeV$^2$, 
with deuteron binding energy $\epsilon_b \approx 2.2$~MeV. 
Another advantage is the positiveness of $t = (p_p-p_D)^2 >0$; thus no zero 
crossing issues arise. These features make the extrapolation procedure in 
reactions (\ref{semidis}) very precise. Because of this, the pole extrapolation 
in processes involving the deuteron is considered as a main method in 
extraction of  different neutron structure functions at future electron-light-ion 
colliders~\cite{CWeiss}.

\section{Theoretical Framework of Tagged Spectator DIS}
\label{sec3}
From 
the above discussion, it follows that  reaction (\ref{semidis}) is 
well suited  for the extraction of neutron structure functions using  the pole 
extrapolation method. 
The reaction (\ref{semidis}) can be described through four nuclear structure 
functions $F^{\text{SI}}_{L,T,TL,TT}$, which depend on $Q^2,x,\alpha_s, \bm 
p_{s\perp}$, where $\mathbf{p}_s$ is the proton momentum, $\alpha_{s} = 
2\frac{E_{s}-p^z_s}{E_D-p^z_D}$ is the light-cone momentum fraction of the 
deuteron carried by the spectator proton normalized such that  $\alpha_s + 
\alpha_i = 2$ ($\alpha_i$ is the 
equivalent quantity for the struck neutron).  The virtual photon has  energy 
$\nu$ and momentum $\bf q$, $Q^2={\bf q}^2-\nu^2$, Bjorken $x={Q^2\over 
2m_N\nu}$, and $\hat z|| {\bf q}$.  
Considering the proton integrated over the azimuthal angle $\phi$ in the 
laboratory
frame, one obtains: 
 \begin{eqnarray}
&&{d\sigma\over dx dQ^2 d^3p_s/E_s} = {4\pi\alpha_{\text{EM}}^2\over x Q^4}
\left(1-y-{x^2y^2m_N^2\over Q^2}\right) \nonumber \\
&&\times\left[F^{\text{SI}}_{2D}(Q^2,x,\alpha_s, \bm p_{s\perp}) + {2\nu 
\tan^2{\theta\over 
2}\over m_N}F^{\text{SI}}_{1D}(Q^2,x,\alpha_s, \bm p_{s\perp})\right],\nonumber 
\\
\label{SI}\label{eq:cross}
\end{eqnarray}
where 
$F^{\text{SI}}_{2D} = F^{\text{SI}}_{L} + \frac{Q^2}{2q^2}\frac{\nu}{ 
m_N}F^{\text{SI}}_T$, 
$F^{\text{SI}}_{1D}= \frac{F^{\text{SI}}_{T}}{ 2}$, and
$y = \frac{\nu}{ E_e}$.   

\begin{figure}[ht]
\centering\includegraphics[width=0.4\textwidth]{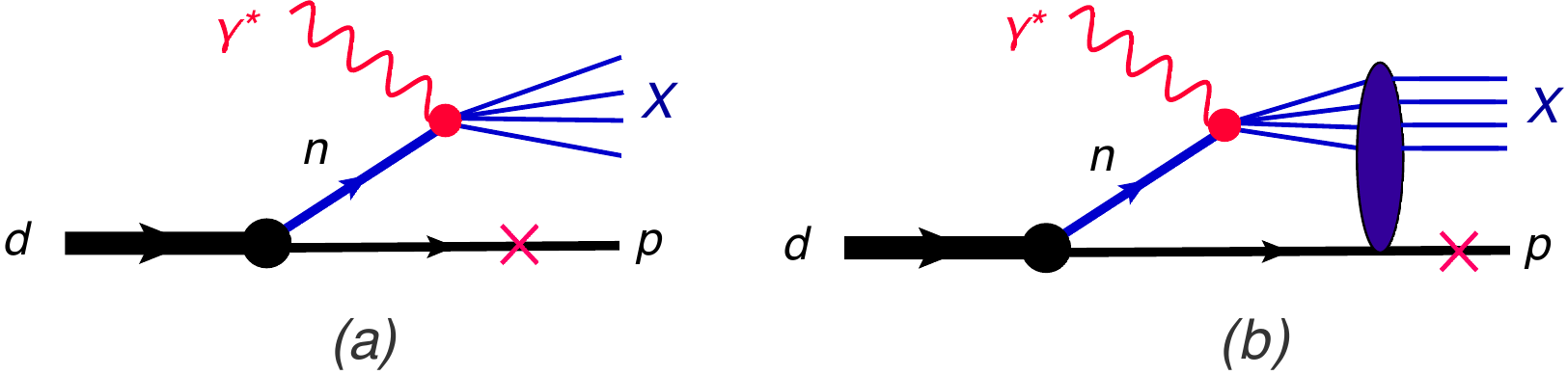}
\caption{(Color online) IA(a) and FSI(b) contributions to reaction 
(\ref{semidis}).}
\label{fig:pwia_fsi}
\end{figure}

The calculation of Eq.~(\ref{eq:cross}) at  
$p_s< 700$~MeV/c  and $x> 0.1$ is based on the 
assumption~\cite{MSS97,polext06,CS11}
that the scattering   proceeds through the interaction of 
the virtual photon off one of the bound nucleons in the deuteron.
Two main diagrams  contribute: IA~[Fig.~\ref{fig:pwia_fsi}(a)] and
final-state interaction~(FSI) diagrams  [Fig.~\ref{fig:pwia_fsi}(b)],
where the latter accounts for the rescattering of the recoil nucleon off the 
products of DIS. While the calculation of the IA term requires the knowledge of 
the deuteron wave function and the treatment of the off-shellness of the bound 
nucleon, the FSI term requires in addition the modeling of the 
deep inelastic  rescattering dynamics.
In Ref.~\cite{CS11}, we developed  a theoretical  model for the calculation of 
the FSI contribution based on the 
extension of the generalized eikonal approximation (GEA) model~\cite{gea,ms01}
to the DIS  domain (see also Ref.~\cite{CMS}). The off-shell 
effects were treated within the 
virtual nucleon approximation (VNA), which works reasonably well for up to 
$\ ~\sim 500$~MeV/c of 
spectator  nucleon momenta, as our previous experience in 
the quasi-elastic regime shows~\cite{noredepn,dreview,Boeglin11}.

The calculations based on this approach\cite{CS11} demonstrated a good agreement with the first data from 
Jefferson Laboratory~(JLab)~\cite{deeps} (referred to as {\em Deeps} data) in 
the $x> 0.3$ and $p_s\ge 300$~MeV/c 
region, describing all the major features of the angular and momentum 
distributions. 
The conclusion from these comparisons was that at $x>0.3$, the FSI is dominated by
the so-called compound  DIS products scattering off the recoil nucleon, which 
 can be 
characterized by a diffractive scattering amplitude.
The wide kinematical range of Ref.~\cite{deeps} allowed us to extract the deep inelastic
FSI cross sections as a function of $x$ and $Q^2$. 
The success in the description of the data~\cite{deeps} motivated 
us to apply the pole extrapolation to extract the neutron 
structure functions $F_{2n}(x,Q^2)$.  The data, however, were taken  at 
large recoil momenta  $p_s \ge 300$~MeV/c, rendering large uncertainties in 
the pole extrapolation  procedure~\cite{polext11}.
More recently, a dedicated tagged DIS experiment  was completed   by the 
BONuS Collaboration~\cite{bonus1,bonus2}, where the recoil  proton  was 
measured at unprecedentedly small momenta of $78$~MeV/c.
These data are the first in their kind for which the pole extrapolation can be 
performed with a higher degree of  accuracy.

\section{Pole Extrapolation of Tagged DIS Processes}
\label{sec4}
The  pole extrapolation in (\ref{semidis}) uses the fact the 
IA amplitude of Fig.~\ref{fig:pwia_fsi}(a) 
[similar to  Eq.~(\ref{IA})]  can be expressed as~\cite{polext06}
\begin{eqnarray}
M^\mu_{IA} =  \langle X|J^{\mu}_{\text{EM}}(Q^2,x)|n\rangle\bar u(p_d - 
p_s)\bar 
u(p_s)\nonumber \\
 \frac{\Gamma_{d\to pn}\chi_d}{ 
|\epsilon_b|(M_d + m_n - m_s) + 2 M_d T_s},
\label{A_IA2}
\end{eqnarray}
where $T_s$ is the kinetic energy of the spectator proton. 
In Eq.~(\ref{A_IA2})
the  pole is associated  with negative kinetic energy of the spectator at:
$T_s^{\text{pole}} = -{|\epsilon_b|\over 2}(1 + {m_n-m_p\over M_d}) \approx 
-{|\epsilon_b|\over 2}$.
While the above IA amplitude diverges at $T_s\rightarrow 
T_s^{\text{pole}}$, the FSI amplitude is 
finite due to an extra loop integration.
In the  
$T_s\rightarrow T_{s}^{\text{pole}}$ limit~\cite{polext06}
\begin{eqnarray}
 M^\mu_{FSI}  & \rightarrow & 
J^{\mu}_{\text{EM}}(Q^2,x)
\bar u(p_d-p_s)\bar u(p_{s})\Gamma_{d\rightarrow pn}\chi_{d}\nonumber \\
& &  \int{d^3k\over 2k^2 (2\pi)^3} {A_{FSI}(k)\over 
2(m_N+T^{\text{pole}}_s-k_0)},
\label{A_FSI2}
\end{eqnarray}
where $k_0= E_s - \sqrt{m_p^2 + (p_s-k)^2}$ and $A_{FSI}$ is the diffractive-like 
amplitude of the rescattering of DIS  products off the 
spectator proton. 
Equation~(\ref{A_FSI2}) is finite at  
the pole as compared to the singular behavior of IA term. 
This result is the essence of the so-called loop theorem~\cite{polext06}.

The pole extrapolation procedure for the extraction of $F_{2n}$ consists 
of multiplying 
the measured structure function, 
$F^{SI,\text{EXP}}_{2D}$[Eq.~(\ref{eq:cross})]  by the factor 
$I(\alpha_s,\bm p_{s\perp},t)$~\cite{polext06}, which 
cancels the singularity of the IA amplitude and is normalized such that
\begin{equation}
F^{\text{extr}}_{2n}(Q^2,x,t) = I(\alpha_s, \bm p_{s\perp},t)
F^{SI,\text{EXP}}_{2D}(Q^2,x,\alpha_s, \bm p_{s\perp}) 
\label{f2extr}
\end{equation}
approaches the free $F_{2n}(Q^2,x,t)$ in the  
$t\rightarrow m_n^2$  limit with FSI effects being diminished.

\section{Pole Extrapolation of the BONuS  Data}  
\label{sec5}

We applied the above described method to the BONuS data~\cite{bonus1,bonus2}, 
which covers  the kinematic range of  
$0.93\le Q^2   \le 3.38$~GeV$^2$ and 
invariant mass of the DIS products 
$ 1.18\le W \le 2.44$~GeV.
The spectator proton was detected at $p_s = 77.5$, $92.5$, $110$, $135$ MeV/c, 
covering a wide angular range of  $-0.9 \le \cos\theta_{s} \le 0.9$.  

\subsection{Renormalization procedure:}
The problem in implementing the pole  extrapolation procedure directly was 
in the fact that in the BONuS experiment different spectator momenta were 
measured at different and poorly known efficiencies.  In the  BONuS analysis  
this issue was solved by normalizing the data for each $p_s$
bin at  $\cos\theta_s\leq -0.2$ to an IA model~\cite{bonus2}.
For our analysis, we chose to 
renormalize the data to the  VNA calculation~\cite{CS11} discussed in Sec.~\ref{sec3}, 
since it also contains  the FSI effects (referred in the text as VNA FSI).
For these calculations  we used the parameterization of the FSI amplitude 
($A_{FSI}$)  obtained 
from the comparisons with the {\em Deeps} data~\cite{CS11}.

\begin{figure*}[ht]
\centering\includegraphics[width=\textwidth]{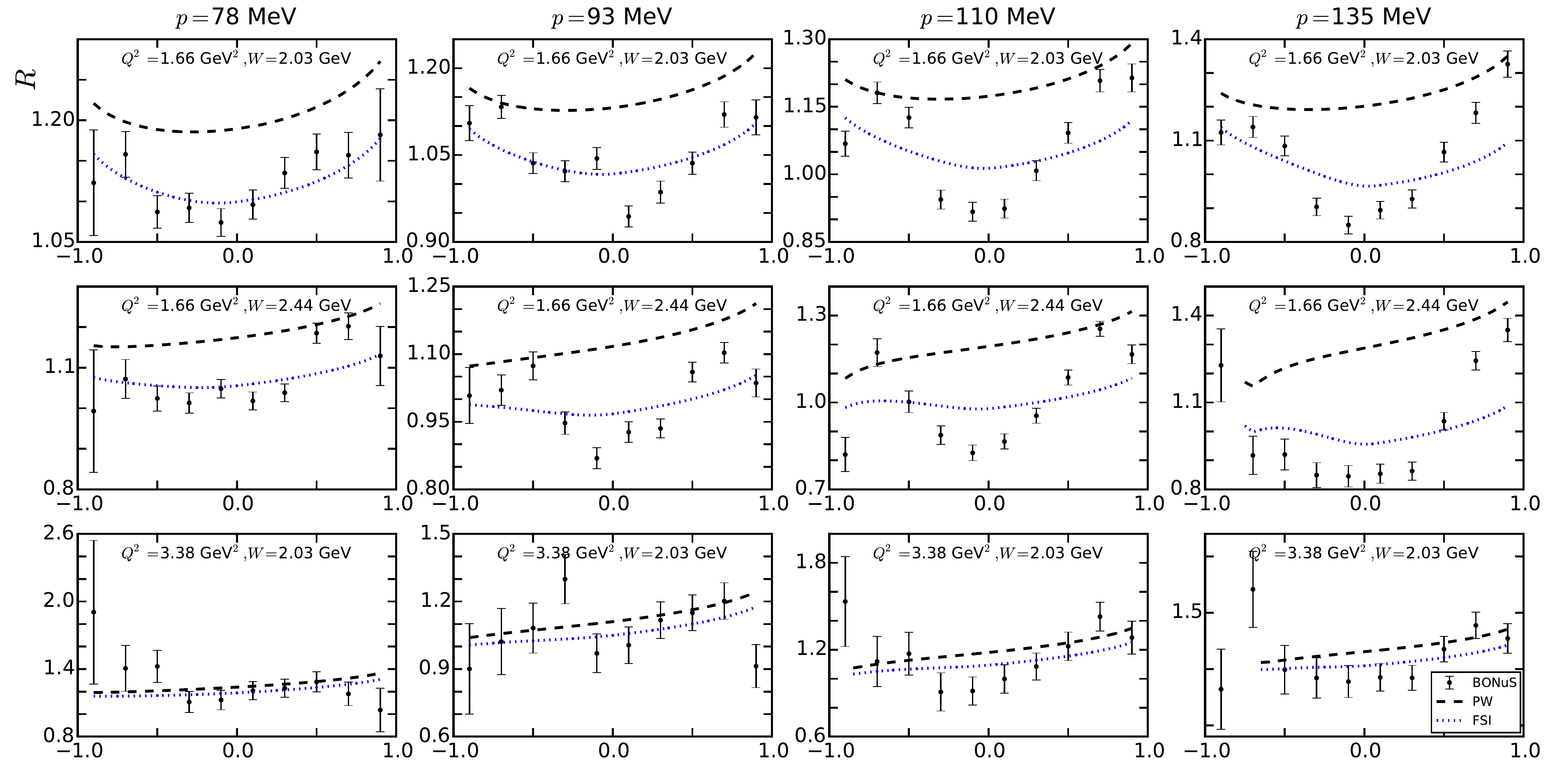}
\caption{(Color online)  Ratio $R$ of the BONuS data to a plane-wave model 
(see Ref.~\cite{bonus2} for details) as a function of spectator 
$\cos\theta_s$ 
compared to our VNA IA (black dashed 
curve) and FSI (dotted blue curve) 
calculation for $E_{\text{beam}}=4.23$~GeV.  An overall 
normalization factor was fitted for each $p_s$ value in the model to 
the FSI calculation; see text for details.  The IA calculation is shown using 
the same normalization factor.}
\label{fig:bonus1}
\end{figure*}

\begin{figure*}[ht]
\centering\includegraphics[width=\textwidth]{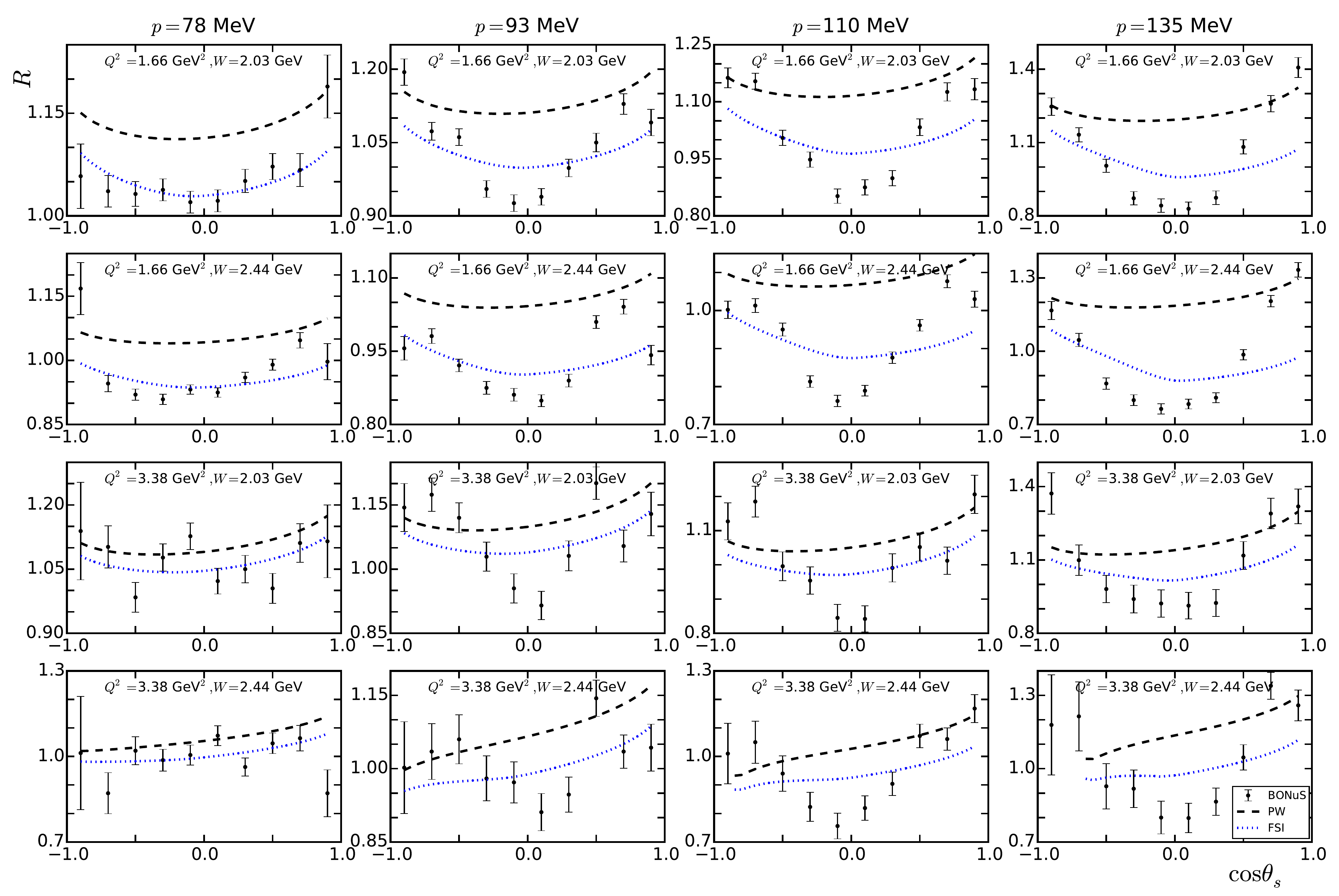}
\caption{(Color online) As Fig.~\ref{fig:bonus1} but for 
$E_{\text{beam}}=5.27$~GeV.}
\label{fig:bonus2}
\end{figure*}

The BONuS data are presented as ratios $R$ of the 
BONuS data to the specific  plane-wave impulse approximation (BONuS IA) model 
discussed in Ref.~\cite{bonus2}. The overall normalization of 
the data was fitted for each spectator momentum setting for two values of initial beam energies.
To obtain the absolute cross sections (required for the pole extrapolation), 
we first multiplied the reported $R$ ratios 
 by the  BONuS IA calculation.   
Then these cross sections   have been fitted to  our VNA FSI calculations
for  each  experimental $p_s$ setting and initial beam energy in the range 
of $x<0.5$ where  neutron DIS structure functions 
are sufficiently well known and have small contributions from nuclear effects.
The results are presented in 
Figs.~\ref{fig:bonus1}~and~\ref{fig:bonus2}.  
For each column in these figures,
corresponding  to a fixed 
spectator momentum $p_s$ and initial 
electron energy $E_{\text{beam}}$, one overall normalization factor was 
obtained.  The values of the normalization factors with 
their errors are shown in Table~\ref{table:norms}.  
The quoted normalization parameters are relative to the BONuS 
normalization  values.  Since BONuS obtained their absolute 
cross sections by fitting their IA model, we basically renormalized the BONUS cross sections to take into account 
the FSI effects. As was mentioned above, the parameters of FSI are fixed 
from the analysis of the only
existing ({\em Deeps}) experiment of reaction (1)~\cite{deeps}. This experiment 
covered larger values of the spectator proton momentum ($\geq 300 $MeV/c), 
which was good for the extraction of
FSI parameters. However, the restricted kinematics of {\em Deeps} measurements 
prevent us
infrom obtaining the sufficiently detailed FSI parametrization 
to be able to
describe the shape of the ratio R in a more refined way in the
$\cos(\theta_s) \approx 0$ region
dominated by the FSI of produced resonances. Further refinements of FSI 
parameters will
allow us to address the more detailed structure of the cross section dominated 
by resonance
production.

In Figs.~\ref{fig:bonus1}~and~\ref{fig:bonus2} we also compare 
the results  of 
VNA  calculation within the impulse approximation
(referred to as VNA IA) with the same normalization factors of 
Table~\ref{table:norms}.  These calculations  indicate  that  the  FSI effects 
are not negligible and they 
increase with the spectator momentum $p_s$.

\begin{table}[htb]
\begin{tabular}{ |l|l|l|l|l| }
\hline
$E_{\text{beam}}$&$p_s$& Norm.& Error & 
$\chi^2/\text{dof}$\\
(GeV)&(MeV)&factor&&\\
\hline\hline
\multirow{4}{*}{4.23} & 77.5 & 1.316 & 0.036 & 2.39 \\
& 92.5 & 1.279 & 0.033&4.61\\
& 110 &1.378 &0.041 &11.1\\
& 135 & 1.494& 0.055&10.2\\
\hline
\multirow{4}{*}{5.27} & 77.5 & 1.176 & 0.025 & 6.54 \\
& 92.5 & 1.203& 0.026&12.4\\
& 110 & 1.244&0.031 &15.7\\
& 135 & 1.417&0.047 &20.8\\
\hline\hline
\end{tabular}
\caption{Normalization factors obtained by fitting the BONuS 
data to our VNA model calculations including FSI.}\label{table:norms}
\end{table}


To estimate the errors in the normalization factors 
we investigated their  dependence on several 
VNA  FSI model ingredients: The first is the choice of the deuteron wave function.
To estimate the  wave function uncertainty we used a selection of 
different deuteron wave functions,  AV18~\cite{AV18}, 
CDBonn~\cite{Machleidt:2000ge}, Paris~\cite{Lacombe:1980dr}, and 
WJC1~\cite{Gross:2008ps}, resulting in  $<0.5$\%
variations.  The uncertainty in  the parametrization of the FSI  amplitude 
obtained in 
Ref.~\cite{CS11} resulted in 2--3\% variations.
Finally, 
in the VNA calculation at $x<0.5$, we used 
the neutron structure functions paramterization from 
Ref.~\cite{CBneutron},
whose average  accuracy in this region is  $\sim$3\%. 
  
\begin{figure}[ht]
\hspace{-0.2cm}
\centering\includegraphics[width=0.4\textwidth]
{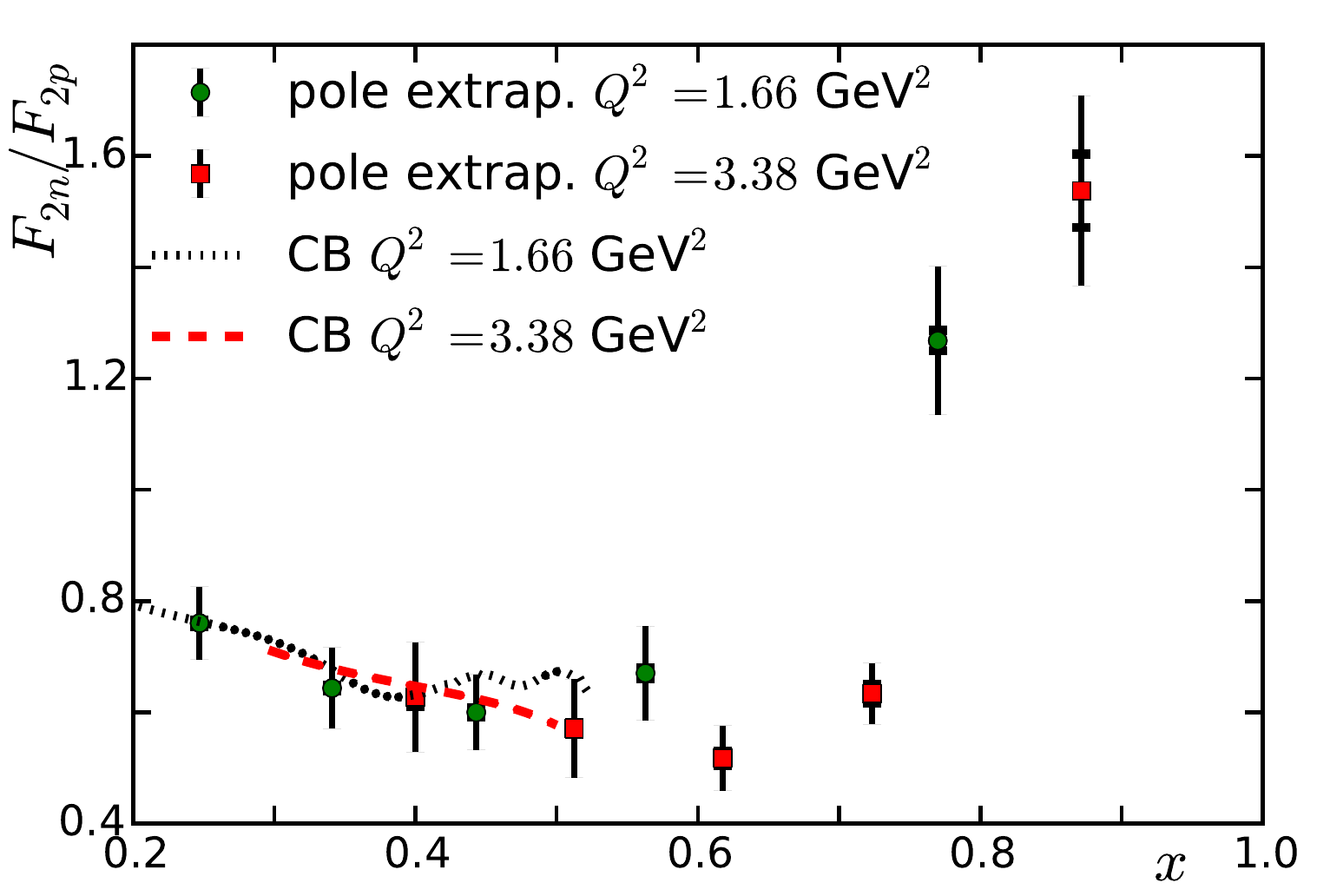}
\hspace{-0.4cm}
\caption{(Color online) $F_{2n}$ to $F_{2p}$ ratio obtained using the pole 
extrapolation applied  to the renormalized BONuS data.  Systematic errors are 
depicted as open error bars.
The dotted black and dashed red curves show the ratio obtained with $F_{2n}$ 
 parametrization of Ref.~\cite{CBneutron}.
The $F_{2p}$  values are estimated  using the fit of Ref.~\cite{CBproton}. } 
\label{fig:f2ratio}
\end{figure}

 As follows from Table~\ref{table:norms}, our fitting procedure yielded 
rather large values for 
 $\chi^2/\text{dof}$. This is mainly due to small values of the statistical 
errors in BONuS data, which were taken 
 into account in the fitting procedure.  Inclusion of the systematic errors of BONuS data  will 
 significantly reduce the magnitude of  $\chi^2/\text{dof}$. However, we did  
not include the latter  into the fit 
 since it was unclear whether these errors are uncorrelated.
 Finally, as a test that  the renormalization procedure yielded valid results, 
we compared our extracted (by pole 
 extrapolation method) 
 neutron structure functions at $x<0.5$ with the data available from the 
analysis of inclusive $d(e,e')X$ reactions~\cite{CBneutron}.
 Note that the inclusive $d(e,e')X$ reactions at  $x<0.5$ have small 
contribution from nuclear effects and neutron data 
 in this kinematics can be considered rather reliable.  As 
Fig.~\ref{fig:f2ratio} shows, our results agree reasonably well with 
 that of Ref.~\cite{CBneutron}.

\subsection{Pole extrapolation in the $x>0.5$ regions}
  
The experimental uncertainties in the absolute cross section of  the BONuS data depend only 
on the momenta  of the spectator proton $p_s$ at given $E_{\text{beam}}$ and 
not on the range of the 
$x$ being probed.
Therefore, the normalization factors
that we obtained for the $x<0.5$ region, can be applied 
to the BONuS data set also in the $x>0.5$ region, where the neutron structure 
functions are not well constrained.
In this way, we obtained the renormalization of the BONuS data set for the 
whole 
measured region of $x$. Using these  renormalized data in 
the pole extrapolation procedure described in Sec.~\ref{sec4}, we extracted 
the neutron structure functions for 
the whole range of the $x$ covered by BONuS experiment.

The extracted $F_{2n}$  for the largest   two $Q^2$ bins of the BONuS 
experiment 
are presented in Fig.~{\ref{fig:f2ratio} and Table~\ref{table:f2ratio}.
These represent  the weighted average of the 
extrapolated values taken over all $\theta_s$ bins. 
The final statistical errors are  similar to those of 
the BONuS data averaged over backward $\theta_s$~\cite{bonus1}.
 
Our procedure of renormalization and pole extrapolation rendered 
systematic errors, which are presented (as open error bars) in 
Fig.~\ref{fig:f2ratio}.  To estimate the systematic errors in the extracted 
$F_{2n}$ values, we took into account statistical and systematic errors of the 
BONuS non-normalized cross sections as well as errors in the estimation of 
the renormalization coefficients of Table~\ref{table:norms}.

To estimate the final systematic errors in the $F_{2n}$ which are  extracted 
in the  pole extrapolation,
we carried out a Monte Carlo simulation where  
each of the inputs in the pole extrapolation (BONuS un-normalized cross sections,  
renormalization coefficients and the uncertainty in the factor $I(\alpha_s,p_{s\perp},t)$ 
due to the choice of the particular deuteron wave function) are distributed randomly in the  Gaussian form 
with a width  corresponding to their estimated errors.  With such randomly 
distributed 
input values the pole extrapolation is  carried out. The widths of the 
distribution 
of the extracted  $F_{2n}$ values are  taken as an estimate for the systematic 
errors in our procedure.  These are the systematic errors quoted in 
Table~\ref{table:f2ratio}. Note that in the future experiments~\cite{newBonus} 
these  systematic errors can be largely reduced by achieving  more reliable 
absolute measurements of the data.
 
\begin{table}[htb]
\begin{tabular}{ cccccccc }
\hline\hline
$Q^2$(GeV$^2$)&$x$&$F_{2n}$&Stat.&Sys.&$\frac{F_{2n}}{F_{2p}}$&Stat.&Sys.\\
\hline
\multirow{5}{*}{1.66}&0.25&0.251&0.002&0.022&0.761&0.006&0.066\\
&0.34&0.181&0.002&0.020&0.644&0.006&0.073\\
&0.44&0.153&0.002&0.017&0.600&0.008&0.067\\
&0.56&0.118&0.002&0.015&0.671&0.010&0.084\\
&0.77&0.090&0.001&0.009&1.268&0.019&0.132\\
\hline
\multirow{5}{*}{3.38}&0.40&0.147&0.004&0.023&0.628&0.017&0.10\\
&0.51&0.091&0.002&0.014&0.571&0.010&0.089\\
&0.62&0.061&0.001&0.007&0.518&0.013&0.057\\
&0.72&0.046&0.001&0.004&0.634&0.017&0.052\\
&0.87&0.030&0.001&0.003&1.54&0.066&0.158\\
\hline\hline
\end{tabular}

\caption{$F_{2n}$ and its ratio to $F_{2p}$ with  statistical and 
systematic errors, obtained with  
the pole extrapolation method applied  to 
the renormalized BONuS data.}
\label{table:f2ratio}
\end{table}

\section{Discussion of the Results} 
\label{sec6}

 The most important advantage of the pole extrapolation method is that the 
extracted neutron structure functions are free from Fermi motion and nuclear medium 
modification effects  which are the main 
and unresolved issues in high-$x$  extractions
in inclusive DIS off the deuteron.  
Our results for $F_{2n}/F_{2p}$  at $x<0.5$ are in fair agreement with the neutron structure functions extracted 
from the analysis of the inclusive data where no significant nuclear 
effects are expected. However our results
exhibit a few surprises at larger $x$~(Fig.~\ref{fig:f2ratio}).  
First, at $x>0.6$,  $F_{2n}$  is larger than the  one extracted in inclusive 
DIS. Note,  however that Fermi effect uncertainties in inclusive DIS 
analyses~\cite{Arrington} still allow the 
values obtained in Fig.~\ref{fig:f2ratio}.
The second interesting property of  our results is the  weak slope  of the 
$F_{2n}/F_{2p}$ ratio with increasing $x$, even indicating 
a possible upward turn of the ratio at $x\gtrsim 0.7$.  
The upward turn is observed also  in the $d(e,e^\prime)X$  
analysis~\cite{src_emc} for up to $x=0.7$, in which the medium modification 
effects in the deuteron are estimated using the 
observed correlation between nuclear EMC and short-range correlation effects. 

Our analysis was applied to the data  beyond 
$x=0.7$ and the intriguing  result is that 
the tendency of the $F_{2n}/F_{2p}$ ratio to increase continues.
It is worth mentioning that the extracted slope  of  $F_{2n}/F_{2p}$  is 
nearly insensitive to our normalization procedure.
Due to  sub-DIS values  of  $W$ ($\approx 1.18~\text{GeV}$) corresponding to 
the highest $x$ values  in Fig.~\ref{fig:f2ratio} one can not directly relate 
the rise of $F_{2n}/F_{2p}$ to   underlying properties 
of the $u$- and $d$-quark distributions at $x\rightarrow 1$.  
At these $W$ such a rise 
is related to the $F_{2n}$ of the $\Delta$ production.  Thus the  relation to the properties of 
quark distributions can be made only based on duality arguments. 
It is worth mentioning that  the recent duality 
paper~\cite{DualityBonus} analyzing the same BONuS data 
concluded that  the $\Delta$-resonance contributes to the duality, 
within 20-30\% accuracy.

If one assumes, however, that the observed  $F_{2n}/F_{2p}$  rise will persist 
in the true DIS region 
then it is intriguing  that  such a  rise   can be an 
indication of the  existence of an isosinglet 
$qq$ short-range correlations~(SRCs) in the nucleon at $x\rightarrow 1$.  Such a 
correlation will 
result in the same momentum sharing effect, 
which  is observed  recently  in asymmetric nuclei in the $NN$ SRC
region~\cite{newprops,scipaper}.  According to this observation, 
the SRC between unlike 
components in the asymmetric  two-Fermi system will result in the small 
component's  dominance in the correlation region such 
that
\begin{equation}
f_1 n_1({p}) \approx f_2 n_2({p}),
\end{equation}
where $f_i$ are the fractions of the components and $n_i(p)$ the high momentum 
distributions normalized to unity.
If such a  $qq$  SRC  would be present in the nucleon, then the above 
equation will translate to 
\begin{equation}
u(x) \approx d(x)
\end{equation}
at $x\rightarrow 1$, since the valence $u$ and $d$ quarks are 
normalized to their respective fractions. Such a relation 
will result in the the rise of the $F_{2n}/F_{2p}$ ratio in the region of $x$ in 
which the $qq$ correlations are dominant.
Note that the possible dominance of  isosinglet $ud$ SRCs
is consistent with the 
flavor decomposition of neutron and proton form factors in the large $Q^2$ 
region~\cite{flavorFF}.

\section{Conclusion and Outlook:} 
\label{sec7}

For the first time the pole extrapolation 
  procedure is used to extract $F_{2n}$ 
from semi-inclusive scattering from the deuteron with a tagged recoil proton.  The 
extracted results are free from Fermi and medium 
modification effects.  They  indicate a possible inversion of the 
decrease of the $F_{2n}/F_{2p}$ ratios at large $x$.
If such an increase would observed in the true 
DIS region, it suggests the  dominance of 
a short-range isosinglet $ud$ correlations, which will result in the momentum 
sharing effects predicted for  asymmetric two-component Fermi systems in which a short 
interaction takes place between  unlike components.\\

\medskip

\noindent {\bf Acknowledgments:} We are thankful to O.~Hen, S. Kuhn, 
E.~Piasetzky 
S.~Tkachenko, 
M.~Strikman and Ch.~Weiss
for helpful discussions.
The work is supported by 
Research Foundation Flanders
and U.S. DOE grant under Contract No. DE-FG02-01ER41172. 
The computational resources used in this work were 
provided by
Ghent University, the Hercules Foundation, and the Flemish government.
We are thankful also to the Jefferson Lab theory group for support and 
hospitality, 
where part of the research has been conducted.

\end{document}